\documentclass[12pt]{iopart}
\usepackage[english]{babel}
\usepackage[latin1]{inputenc}
\usepackage{hyperref}

\usepackage{graphicx}
\graphicspath{{images/}}
\usepackage{booktabs}
\usepackage{multirow}

\usepackage{amsmath}
\usepackage{xcolor}
\usepackage{mathrsfs}
\usepackage{iopams} 

\def\pmx{\begin{pmatrix}}
\def\emx{\end{pmatrix}}

\newcommand{\ket}[1]{|#1\rangle}
\newcommand{\bra}[1]{ \langle #1 \,  |}

\DeclareMathOperator{\eins}{ {\bf 1}}

\begin{document} 

\title{Hypergraph states in Grover's quantum search algorithm}

\author{M. Rossi$^1$ $^\ast$, D. Bru{\ss}$^2$ and C. Macchiavello$^1$}

\address{$^1$Dipartimento di Fisica and INFN-Sezione di Pavia,
Via Bassi 6, 27100 Pavia, Italy\\
$^2$Institut f{\"u}r Theoretische Physik III, 
Heinrich-Heine-Universit{\"a}t D{\"u}sseldorf, D-40225 D{\"u}sseldorf, 
Germany}

\ead{$^\ast$ matteo.rossi@unipv.it}

\begin{abstract}
We study the entanglement content of the states employed in the Grover 
algorithm after the first oracle call when a few searched items are 
concerned. We then construct a link between these initial states
and hypergraphs, which provides an illustration of their entanglement properties. 
\end{abstract}

\date{\today}

\section{Introduction}

Even though entanglement is considered as a major resource in quantum 
information processing, the role it plays in achieving the quantum 
computational 
speed-up in the currently known quantum algorithms still remains to be elucidated.
In Ref. \cite{jl} it was shown that in  Shor's algorithm  multipartite quantum
entanglement is needed
to attain exponential computational speed-up.
The presence of multipartite 
entanglement in the Deutsch-Jozsa algorithm and in the initial
step of the Grover algorithm was pointed out recently \cite{ent-algo}. Moreover, multipartite 
entanglement was shown to be present at each computational step in Grover's 
algorithm and a scale invariance property of entanglement dynamics
was proved \cite{grover-lett}.
Lately, the notion of quantum hypergraph states was put forward
and their link to states employed in the Deutsch-Jozsa and
Grover algorithms was proved \cite{hyper}. In this paper we
consider the initial states employed in the Grover algorithm 
for a small number of searched solutions, study their
entanglement content (in terms of the geometric measure of entanglement)
 and show the explicit connection to hypergraph
states.

The paper is organised as follows. In Sect. \ref{s2} we review the 
notion of multi-qubit real equally weighted states and their link to 
hypergraphs. In Sect. \ref{s3} we consider the Grover algorithm and analyse 
the entanglement content of the corresponding states in the initial step 
of the algorithm for different numbers of solutions. 
We derive the hypergraphs underlying these symmetric initial states 
for one and two solutions in Sect. \ref{s4}, and we finally summarise the main results in 
Sect. \ref{conc}.

\section{Real equally weighted states and hypergraphs}
\label{s2}

The $n$-qubit register employed in the Deutsch-Jozsa \cite{dj} and in 
the Grover \cite{grover} algorithms is initially prepared in state
\begin{equation}
\ket{\psi_0}\equiv \frac{1}{\sqrt{2^n}}\sum_{x=0}^{2^n-1}\ket{x}\;,
\label{psi0}
\end{equation}
which corresponds to the equally weighted 
superposition of all possible $2^n$ states $\ket{x}$ 
in the computational basis.
The next step in both algorithms consists in applying 
a unitary transformation $U_f$ which generates the state
\begin{equation}
\ket{\psi_f}\equiv \frac{1}{\sqrt{2^n}}\sum_{x=0}^{2^n-1}(-1)^{f(x)}\ket{x}\;,
\label{psif}
\end{equation}
where $f(x)$ is the $\{0,1\}^{n}\to \{0,1\}$ Boolean function that needs to be 
evaluated in the considered algorithm. 
Notice that $(-1)^{f(x)}=\pm 1$ is just a real phase factor.
The above states are referred to as multi-qubit ``real equally weighted 
states'' (REW) $\ket{\psi_f}$. 

The REW states were recently linked to hypergraphs and the set of REW states 
was proved to be the same as the one of quantum hypergraph states \cite{hyper}.
A quantum hypergraph state is defined as follows.
A hypergraph $g_{\leq n}=\{V,E\}$ is a set of $n$ vertices $V$ with a set of  
hyperedges $E$ of any order $k$ ranging from $1$ to $n$
(a hyperedge of order $k$ connects a set of $k$ vertices).
Given a mathematical hypergraph, the corresponding quantum state can be found 
by following the three steps:
assign to each vertex a qubit and initialise each qubit as $\ket{+}$ 
(the total initial state is then denoted by $\ket{\psi_0}$).
Wherever there is a hyperedge, perform a controlled-$Z$ operation between 
all  connected qubits. Formally, if the qubits $i_1, i_2, ..., i_k$  are 
connected by a $k$-hyperedge, then perform the operation 
$C^kZ_{i_1i_2...i_k}$. 
The gate $C^kZ_{i_1i_2...i_k}$ introduces a minus sign to the input state 
$\ket{11...1}_{i_1i_2...i_k}$, i.e. 
$C^kZ_{i_1i_2...i_k}\ket{11...1}_{i_1i_2...i_k}=-\ket{11...1}_{i_1i_2...i_k}$, 
and leaves all the other components of the computational basis unchanged.
In this way we get the quantum state
\begin{equation}
\ket{g_{\leq n}}=\prod_{k=1}^n
\prod_{\{i_1,i_2,...,i_k\}\in E}C^kZ_{i_1i_2...i_k}\ket{\psi_0},
\end{equation}
where $\{i_1,i_2,....,i_k\}\in E$ means that 
the $k$ vertices are connected by a $k$-hyperedge. Notice that the product 
concerning the index
$k=1,2,...,n$ accounts for different types of hyperedges in the hypergraph.
We remind the reader that the well known graph states constitute a subset of 
quantum hypergraph states: they correspond to ordinary graphs, namely 
hypergraphs with 
all hyperedges being of order $k=2$. In the following sections we will discuss some 
explicit examples of hypergraph states that appear in Grover's algorithm.

\section{Initial states in Grover's algorithm}
\label{s3}

In this section we focus on the case of Grover's algorithm. 
The state (\ref{psif}) is achieved after the first application of the 
oracle. In this case the function $f$ has output 1 for entries $x$ that 
correspond to solutions of the search problem and output 0 for values of $x$ 
that are not solutions. 
We denote with $M$ the number of solutions, which typically
is much smaller than the total number of entries $2^n$. 

We will now study the entanglement properties of states (\ref{psif})
as functions of the number of qubits $n$ for a fixed small number
of solutions, i.e. $M=1,2$. 
We will quantify the amount of entanglement by the geometric 
measure of entanglement \cite{wg}, which for a pure $n$-partite state 
$\ket{\psi}$ reads
\begin{equation}\label{geom}
E_q(\ket{\psi})=1-\max_{\ket{\phi}\in S_q}|\bra{\psi}\phi\rangle |^2\;,
\end{equation}
where $S_q$ represents the set of $q$-separable states, i.e. states
that are separable with respect to $q$ partitions.
Notice that $E_n$ quantifies the amount of 
entanglement of any kind contained in the global system, i.e. it is 
non-vanishing even for states showing 
entanglement just between two subsystems, while $E_2$ quantifies
genuine multipartite entanglement. In the paper we will 
compute $E_n$.

We first review the case of a single solution to the search problem 
($M=1$), considered in \cite{ent-algo}. 
Without loss of generality, as will be proved later, we consider
the state representing the solution to be invariant under 
any permutation of the $n$ qubits (e.g. $\ket{111...1}$). Therefore, the 
state $\ket{\psi_{M=1}}$ after the oracle call, is also permutation invariant.
We first compute $E_n$ for this set of states, as a function of the number of qubits $n$.
Due to the symmetry property, the search for the maximum in Eq. \eqref{geom}
can be restricted to  symmetric separable states 
$\ket{\phi}^{\otimes n}$ \cite{symm}, so that the 
maximisation involves only the two parameters $\alpha\in [0,\pi]$ and 
$\beta\in[0,2\pi]$ that define the single qubit state 
$\ket{\phi}=\cos{\frac{\alpha}{2}}\ket{0} + e^{i\beta}\sin{\frac{\alpha}{2}}
\ket{1}$. 

The geometric measure of entanglement $E_n$ for $M=1$, i.e. for one solution 
of the search algorithm, thus reads
\begin{equation}\label{geom_n1}
E_n(\ket{\psi_{M=1}})=
1-\max_{\alpha, \beta}\frac{1}{2^n} \Big|\big(\cos\frac{\alpha}{2}+e^{i\beta}
\sin\frac{\alpha}{2}\big)^n-2e^{i\beta}\sin\frac{\alpha}{2}\Big|^2.
\end{equation}
The optimal value of $\beta$ can be shown to be zero by induction over the 
number $n$ of qubits, while the optimal $\alpha$ can be determined by 
explicitly calculating the derivative of the overlap, 
and then finding the root of a polynomial in $t=\tan \frac{\alpha}{2}$.

\begin{figure}[t!]
\centering
\includegraphics[width=.49\columnwidth]{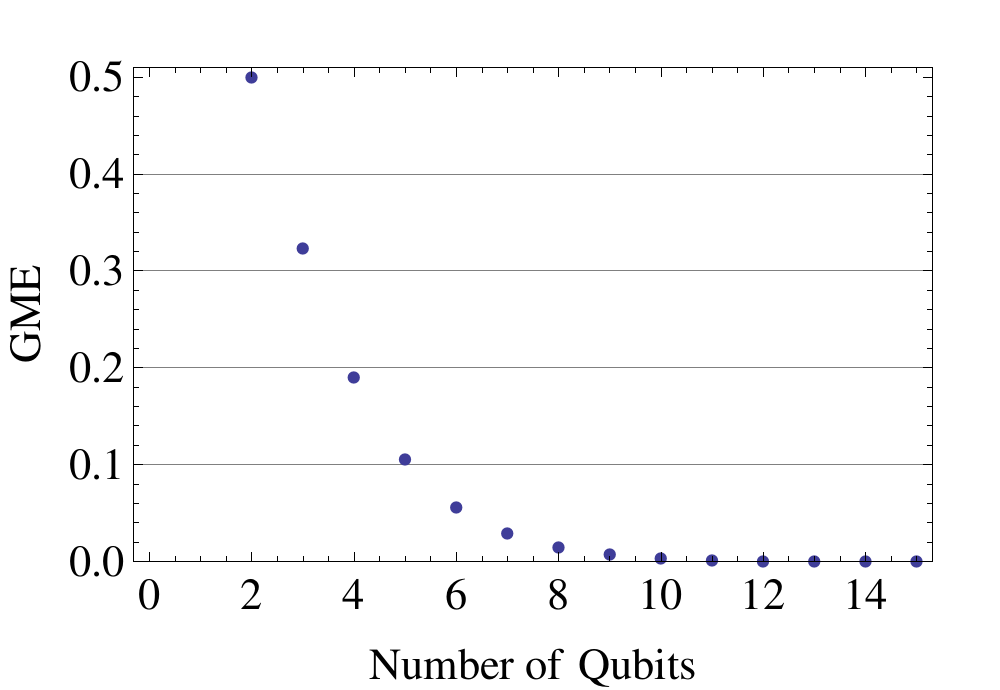}
\includegraphics[width=.49\columnwidth]{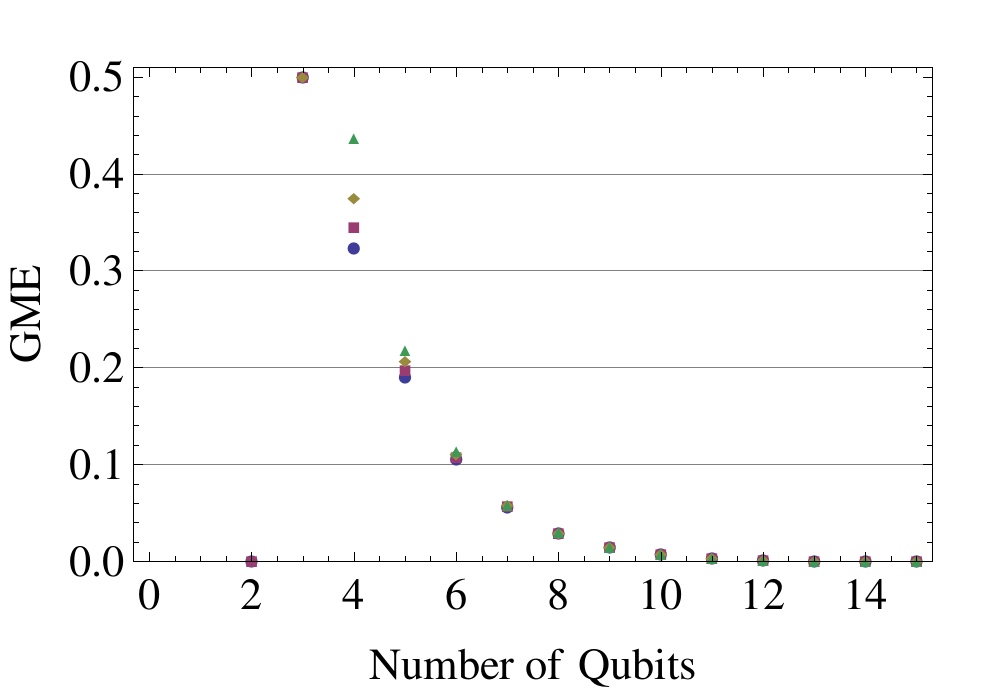}
\caption{Left: Geometric measure of entanglement $E_n(\ket{\psi_{M=1}})$ as a 
function of the number of qubits $n$ for a single searched item. Right: $E_n(\ket{\psi_{M=2}})$ as a function of the number of qubits $n$ for two 
searched items; several values of the Hamming distance $d$ are considered: 
$d=1$ blue dots, $d=2$ purple squares,  $d=3$ yellow diamonds, $d=4$ 
green triangles.}
\label{entM}
\end{figure}

In Fig. \ref{entM} (Left) we show the behaviour of $E_n(\ket{\psi_{M=1}})$,
as a function of the number of qubits $n$.
As we can see, the amount of entanglement decreases for
increasing number of qubits. 
Notice however that, as shown in \cite{grover-lett}, even though the state 
shows an infinitely small amount of entanglement, it is genuine 
multipartite entangled.

We point out that the above results which were explicitly derived for permutation 
invariant states hold also for any Grover search with one searched item. 
Actually, all these states can be achieved from a symmetric state  by applying 
tensor products of $\sigma_x$ Pauli operators and identity 
operators $\eins$ (e.g. $\ket{001...1}=\sigma_{x1}\otimes\sigma_{x2}\otimes
\eins_3....\ket{111...1}$). Since these operations are local, 
they do not change the entanglement content of the resulting state.

We now focus on the states employed in the case of two solutions of the 
search problem ($M=2$), i.e. states \eqref{psif} with two minus signs. 
We introduce a  classification of these states based on the Hamming 
distance $d$ between the two computational basis states representing the 
solutions. We will show that the entanglement properties of 
the regarded states depend crucially on
the number of digits in which the two solutions differ, i.e. on their 
Hamming distance $d$. 

Without loss of generality, as will be proved below, we consider first 
the case in which the two $n$-qubit states representing the solutions differ 
in the first $d$ digits and are invariant under permutations 
of the first $d$ and last $n-d$ qubits, respectively (e.g. 
$\ket{\underbrace{0...0}_d1...1}$ and  $\ket{\underbrace{1...1}_d1...1}$). 
We first compute $E_n$ for this set of states, as a function of $n$.
Due to the permutation invariance property, the search for the closest
separable state, appearing in the geometric 
measure of entanglement,
can still be restricted to separable states that show the same symmetry 
\cite{symm}, i.e. $\ket{\phi}^{\otimes d}\ket{\varphi}^{\otimes n-d}$. 
Therefore the 
maximisation involves only the four parameters $\alpha,\gamma\in [0,\pi]$ and 
$\beta,\delta\in[0,2\pi]$ that define the two single qubit states 
$\ket{\phi}$ and $\ket{\varphi}$.

The geometric measure of entanglement $E_n$ for two solutions with
Hamming distance $d$ then takes the form
\begin{eqnarray}
\label{geom_n2}
E_n(\ket{\psi_{M=2}})=
1-\max_{\alpha,\beta,\gamma,\delta}\frac{1}{2^n} 
\Big|\big(\cos\frac{\alpha}{2}+e^{i\beta}\sin\frac{\alpha}{2}\big)^{d}
\big(\cos\frac{\gamma}{2}+e^{i\delta}\sin\frac{\gamma}{2}\big)^{n-d}\\ 
\nonumber
-2e^{i(n-d)\delta}\sin^{n-d}\frac{\gamma}{2}
		\big(\cos^{d}\frac{\alpha}{2} + 
e^{id\beta}\sin^{d}\frac{\alpha}{2}\big)\Big|^2.
\end{eqnarray}
Notice that when $d=n$ the maximisation procedure involves only two 
parameters: In this case the state is
completely invariant under any permutation of the qubits, 
and thus only two parameters are needed.

As before, the optimal parameters $\alpha$, $\beta$, $\gamma$ and $\delta$ 
can be computed by maximising the squared overlap numerically. 
The corresponding results are reported in Fig. \ref{entM} (Right), where the 
geometric 
measure of entanglement is plotted versus the number of qubits $n$.  
Notice that the state of two qubits is always separable for 
$M=2$ and that $E_n(\ket{\psi_{M=2}})$ for three qubits collapses to the 
single value $1/2$.
As in the case $M=1$, we can see that 
$E_n$ approaches zero exponentially fast for increasing $n$ . 
This behaviour holds for any finite value of the Hamming distance $d$.
Notice also that for fixed finite number of qubits $n$ the Hamming distance 
plays a crucial role for the 
amount of entanglement, since states with two solutions with higher distance 
$d$ exhibit a higher amount of entanglement.

Finally, notice that all the results presented so far, even if they were 
explicitly derived for partially permutation invariant states, hold for any 
Grover search algorithm with two searched items. Actually, analogously to
the case of a single solution discussed above, also for $M=2$ all these 
states with fixed Hamming distance $d$ can be reached from a partially 
symmetric one by applying tensor products of $\sigma_x$ Pauli operators and 
identity operators $\eins$ and/or
permutations of the $n$ qubits. As for the single solution case,
since these operations are local, 
they do not change the entanglement content of the resulting state.

As mentioned in Sect. \ref{s2} all the above states are REW states and 
therefore they correspond to hypergraphs. In the next section we show how to 
relate states with $M=1,2$ to hypergraphs.

\section{Connection to hypergraph states}
\label{s4}

Quantum hypergraph states, as reviewed in Sect. \ref{s2}, allow us to describe 
the initial states employed in the Grover algorithm in a very convenient way. 
Indeed, entanglement properties and the gates (of the $CZ$ type) that we 
explicitly need in order to generate them from the separable state
$\ket{\psi_0}$ emerge very naturally from their hypergraph structure.

Consider first the initial symmetric state $\ket{\psi_{M=1}}$ 
(with a minus sign in front of the component $\ket{111...1}$). 
It is straightforward to see that it corresponds to the hypergraph with the 
unique hyperedge of order $n$. Therefore, it has a very simple structure in 
the light of hypergraphs, showing clearly the presence of multipartite 
entanglement.

In order to discuss the case with two minus signs, we first notice that the hyperedge of order $n$ will never appear now. Hence, even though states $\ket{\psi_{M=2}}$ might have a much more complicated hypergraph structure than $\ket{\psi_{M=1}}$, the gate $C^nZ$ will never be involved. This is because the state $\ket{\psi_{M=2}}$ has  an even  number of minuses, while any product of $C^kZ$ operators involving the gate $C^nZ$ is diagonal in the computational basis with an odd number of minus signs.
We now discuss the general rule to find the hypergraph associated to the 
initial  symmetric state $\ket{\psi_{M=2}}$ with general Hamming distance $d$. 
Let $(|\underbrace{0...0}_d\rangle+|\underbrace{1...1}_d\rangle)|
\underbrace{1...1}_{n-d}\rangle$ be the two states with negative sign, then 
the hypergraph associated to $\ket{\psi_{M=2}}$ can be derived as follows: 
Group the last $n-d$ vertices with a hyperedge of order $n-d$. Then, connect 
the whole group to the remaining $d$ vertices in any possible way, namely by 
exploiting hyperedges of any order greater than $n-d$ except the hyperedge of 
order $n$. The above procedure is needed because, after we have generated 
the desired minus signs in front the two components $\ket{0...01...1}$ and  
$\ket{1...11...1}$, we then have to correct the undesired by-product minuses 
in front
of the other components that contain states $\ket{1}$ for all the last
$n-d$ qubits. 

\begin{figure}[t!]
\centering
\includegraphics[width=.45\columnwidth]{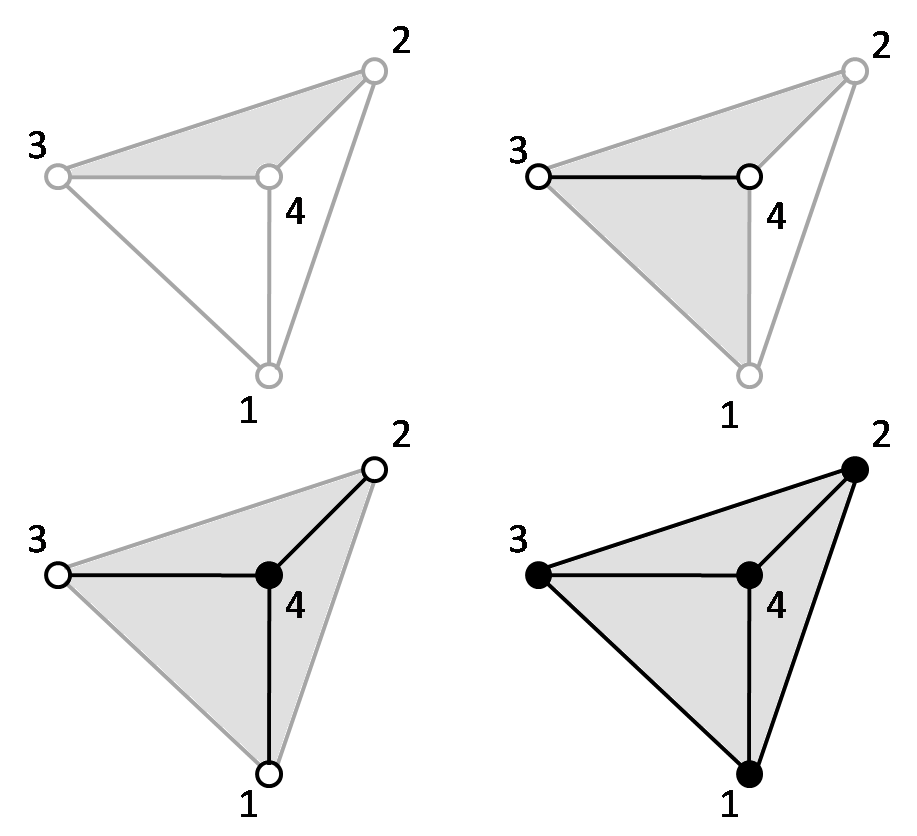}
\caption{Hypergraphs associated to the four-qubit initial Grover states with 
two solutions for several values of the Hamming distance. 
Top-left corner: $d=1$, top-right corner: $d=2$, bottom-left corner: $d=3$, 
and bottom-right corner: $d=4$. Each empty vertex represents a qubit. 
Full dark dots, dark lines and grey faces represent hyperedges of order $1,2$ 
and $3$, respectively. 
Recall that a hyperedge of order $1$ represents a local $Z$ gate.
Notice that in the hypergraph with $d=4$ also the 
hidden face connecting the vertices $1,2$ and $3$ is present.}
\label{hypergrover}
\end{figure}

Notice that both the extreme cases $d=1$ and $d=n$ fit into this scheme. 
Regarding the former, since we are not allowed to draw the hyperedge of 
order $n$, we only group the last $n-1$ vertices without connecting them to 
the remaining one. The biseparability of the state then follows trivially. 
For the latter, we do not apply the first step of the procedure above, 
but we connect all possible vertices according to the second step. 
Notice that, in order to derive the hypergraph corresponding to 
$\ket{\psi_{M=2}}$ with $d=n$, we have to recast $\ket{\psi_{M=2}}$ into the 
hypergraph state with a plus sign in front of $\ket{0...00}$, by multiplying 
all amplitudes by a factor $-1$. As an example, the hypergraphs associated to 
the four-qubit symmetric states $\ket{\psi_{M=2}}$ with Hamming distance 
$d=1,2,3$ and $4$ are shown in Fig. \ref{hypergrover}. 

Notice that the proposed procedure can be generalized to any initial state 
$\ket{\psi_{M=2}}$ by simply renaming and re-ordering the vertices.

\section{Conclusions}
\label{conc}

In this paper we have explicitly computed the amount of entanglement 
contained in the states employed in  the initial step of the Grover 
algorithm, namely after the first oracle call, for one and two solutions. 
Numerical results suggest that the entanglement content of these states, 
quantified by the geometric measure of entanglement, decreases
exponentially  with increasing number of qubits composing the register. 
In the search for two items, the Hamming distance between the items
is found to 
play a crucial role with respect to the entanglement content:
For a fixed number of qubits the state turns out to be more entangled
for a larger Hamming distance.

The connection with quantum hypergraph states has then been worked out, 
showing the mathematical hypergraphs associated to symmetric initial states 
with one or two solutions. We have shown that, besides giving a very convenient pictorial 
representation of quantum states, the hypergraph structure also highlights 
some entanglement properties of the states, such as biseparability or the 
presence of genuine multipartite entanglement.

\end{document}